# Pre-Training of Equivariant Graph Matching Networks with Conformation Flexibility for Drug Binding

*Fang Wu, Shuting Jin, Yinghui Jiang, Xurui Jin, Bowen Tang, Zhangming Niu, Xiangrong Liu, Qiang Zhang, Xiangxiang Zeng, and Stan Z. Li*

The latest biological findings observe that the motionless "lock-and-key" theory is not generally applicable and that changes in atomic sites and binding pose can provide important information for understanding drug binding. However, the computational expenditure limits the growth of protein trajectory-related studies, thus hindering the possibility of supervised learning. A spatial-temporal pre-training method based on the modified equivariant graph matching networks, dubbed ProtMD which has two specially designed self-supervised learning tasks: atom-level prompt-based denoising generative task and conformation-level snapshot ordering task to seize the flexibility information inside molecular dynamics (MD) trajectories with very fine temporal resolutions is presented. The ProtMD can grant the encoder network the capacity to capture the time-dependent geometric mobility of conformations along MD trajectories. Two downstream tasks are chosen to verify the effectiveness of ProtMD through linear detection and task-specific fine-tuning. A huge improvement from current state-of-the-art methods, with a decrease of 4.3% in root mean square error for the binding affinity problem and an average increase of 13.8% in the area under receiver operating characteristic curve and the area under the precision-recall curve for the ligand efficacy problem is observed. The results demonstrate a strong correlation between the magnitude of conformation's motion in the 3D space and the strength with which the ligand binds with its receptor.

## 1. Introduction

The development of a new drug is well known to be very expensive.[1,2] Accurate drug binding prediction is a prerequisite for fast virtual screening,[3,4] which is to understand how drug-like molecules (ligands) interact with the target proteins (receptors). Recently, deep learning (DL)-based methods have emerged to drastically reduce the molecular search space and help accelerate the drug discovery process.[5]

The process of a receptor accommodating a small molecule has been shown to be highly dynamic and time-dependent, thus the initial motionless "lock-and-key" theory of ligand binding[6] has been abandoned. Currently, it is in favor of binding models that account for not only conformational changes, but random dynamic interaction.[7–11] This is because the receptor and ligand flexibility are crucial for correctly predicting drug binding and other related thermodynamic and kinetic properties.[12,13] However, prior DL-based studies concentrate merely on a single, stable, and static conformation[14] without considering the time-dependent mobility.

F. Wu, S. Z. Li
School of Engineering
Westlake University
Hangzhou 310024, China
E-mail: stan.zq.li@westlake.edu.cn
F. Wu, S. Jin, Y. Jiang, X. Jin, B. Tang, Z. Niu
MindRank AI Ltd.
Hangzhou 310000, China

S. Jin, X. Liu
School of Informatics
Xiamen University
Xiamen 361005, China
Q. Zhang
ZJU-Hangzhou Global Scientific and Technological Innovation Center
Hangzhou 311200, China
Q. Zhang
College of Computer Science and Technology
Zhejiang University
Hangzhou 310013, China
X. Zeng
School of Information Science and Engineering
Hunan University
Hunan 410082, China

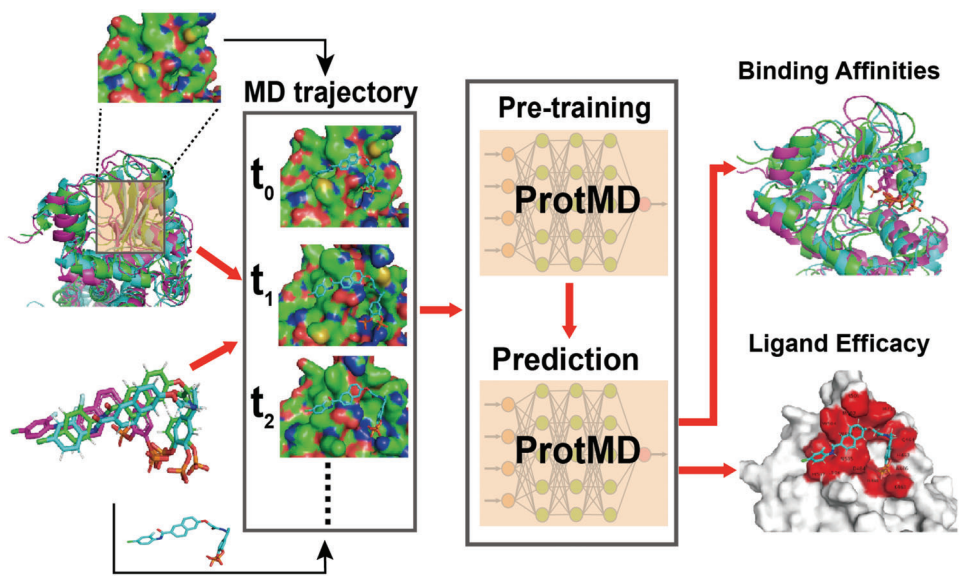

**Figure 1.** High-level overview of our PROTMD based on the MD trajectories. The input during pre-training is a large amount of conformations at different timeframes. The output during inference is a wide variety of drug binding-related properties.

While crystallographic studies have convincingly demonstrated that protein flexibility matters in drug binding, the process demands expensive labor.[15] Alternatively, molecular dynamics (MD) simulations[16] seek to approximate atomic motions by Newtonian physics[17] to reduce needed human labor; also MD can be used to incorporate flexibility into docking calculations. Some MD-based techniques allow a thorough sampling of the conformational space for large biomolecules, and can include the complete description of the pathway of the ligand binding to its target protein.[18–20] However, the cost of MD prevents is prohibitively high with the growing size of reported protein trajectory-related data. Consequently, a completely supervised paradigm of training and inference on MD trajectories of protein-ligand pairs is infeasible.

Motivated by the above-mentioned reasons, in this paper, we aim to explore the mechanism behind the binding prediction from both spatial and temporal perspective, and land on self-supervised learning[21–23] to empower pre-trained models with the ability to learn temporal dependencies. We propose a simple yet effective self-supervised pre-training framework with the full employment of temporal sequences of **Prot**ein structures on **MD** trajectories termed as PROTMD as shown in **Figure 1**. Specifically, two types of self-supervised learning tasks are constructed, one for atom-level and the other for conformation-level, to better capture the internal and global information of MD trajectories. The former is a prompt-based denoising generative prediction, working at the atom-level. It asks the model to produce future conformations based on the current one. Unlike the naive generative self-supervised learning, a time-series prompt is added to regulate and control the time interval between the source and target conformations. This enables the molecular encoder to capture both short-term and long-term dependencies inside the MD trajectories. Apart from that, extra noise is injected into the input conformation to increase the task difficulty to prevent the model from overfitting. This setting conforms to the principles of enhanced sampling mechanism in MD simulations.[24–26] The lat-ter is through a conformation-level snapshot ordering task, which requires the model to identify the temporal order of a set of consecutive snapshots.

To fully unleash the potential of our proposed self-supervised method, we refine an *E(3)-equivariant graph matching network* (EGMN)[27] to cope with ligand binding modeling and use it as the backbone of the PROTMD. The EGMN as geometric network can jointly transforms both the features and 3D coordinates to perform message passing on *intra* and *inter* graphs. In the experiments, we train the PROTMD on the MD trajectories of sixty-four protein-ligand pairs with a total of 62.8K snapshots and then they are leveraged for two downstream drug binding-related tasks, that is, the binding affinity prediction and the ligand efficacy prediction. Our model leads to state-of-the-art results. Clear visualization strongly demonstrate that our pre-training approach can significantly and consistently improve the model performance and learn effective protein representations using MD data. It also proves the extraordinary ability of our PROTMD, pre-trained with a limited number of samples, to generalize to a diversity of downstream tasks. More importantly, we investigate the underlying mechanism behind the success of PROTMD, and further demonstrate a tight correlation between the magnitude of spatial motion of conformation and the extent to which the ligand and the receptor bind with each other. This provides solid evidences that our PROTMD efficiently captures conformations' flexibility during the drug binding process. The source code of this study providing the ProtMD is freely available in "https://github.com/smiles724/ProtMD".

## 2. The Framework of PROTMD

This section gives a brief overview of our PROTMD, highlighted in Figure 1 and details in **Figure 2**. It consists of two parts: various spatial-temporal self-supervised learning tasks and an equivariant graph matching network. See the Experimental Section for more descriptions on PROTMD.

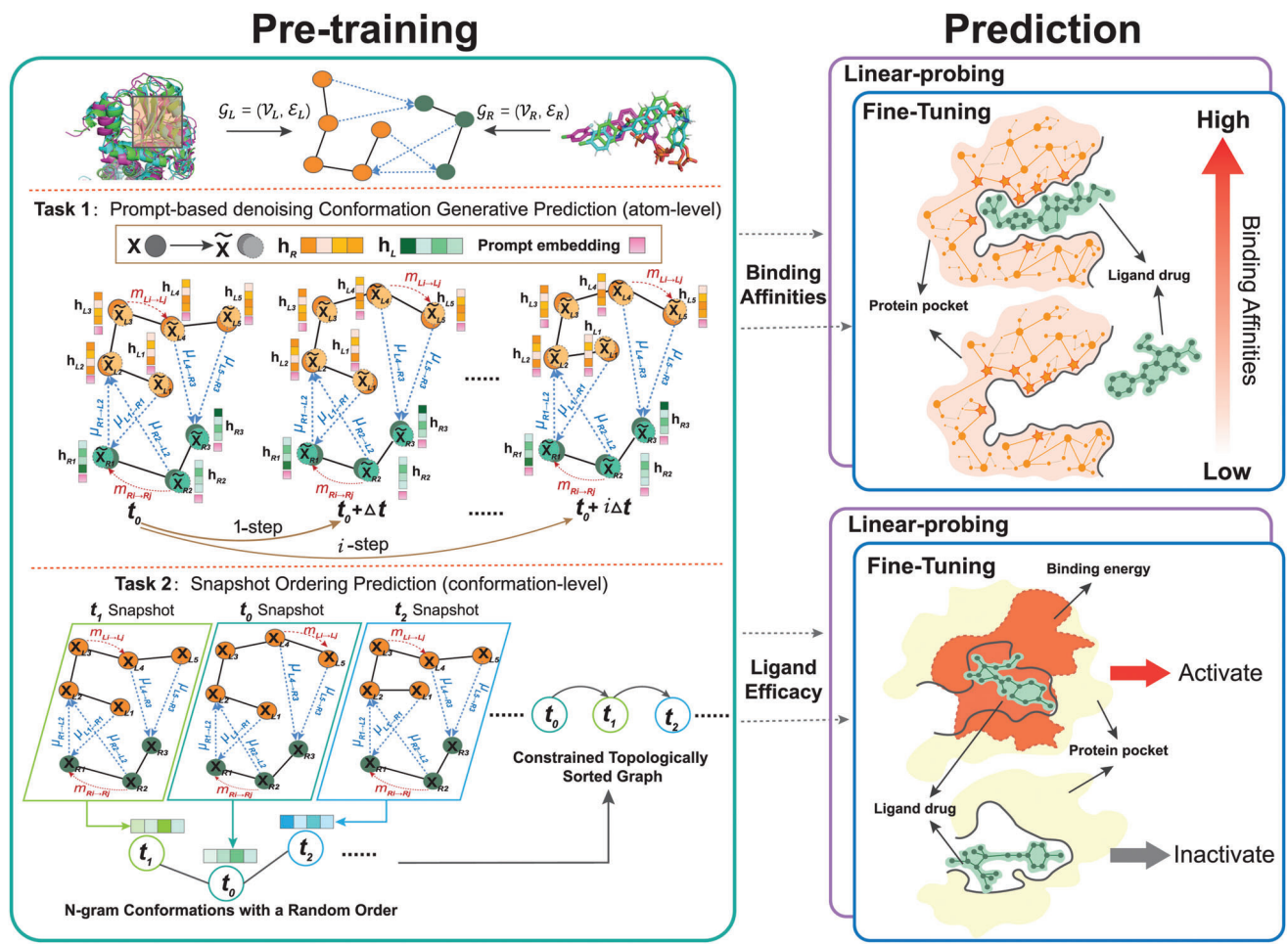

**Figure 2.** Framework of our PROTMD pipelines. In the pre-training stage, two types of self-supervised learning tasks are introduced there to capture drug binding-related information hidden inside the MD trajectories. One is the prompt-based denoising generative prediction task, and the other is the snapshot ordering task. In the fine-tuning and linear-probing stage, the pre-trained model is utilized in two downstream problems, where the binding affinity prediction is a regression problem, and the ligand efficacy prediction is a classification problem.

## 2.1. Spatial-Temporal Self-Supervised Learning Tasks

To capture the temporal information of MD trajectory and boost the generalization ability of the model, we construct a spatial-temporal conformation sequences to represent MD trajectories of protein-ligand pairs and propose two self-supervised learning tasks to pre-train PROTMD model: the prompt-based denoising conformation generative task (atom-level) and the snapshot ordering task (conformation-level). Notably, these two levels of modeling is designed with different purposes. The atom-level task specializes in capturing the local context of each particle, while the conformation-level task is particularly to capture the long-term context within the complexity.

### 2.1.1. Spatial-Temporal Protein Sequences

We consider MD trajectories of each protein-ligand pair with $T$ timesteps. At each time $t \in [T]$, the ligand graph $\mathcal{G}_L^{(t)} = (\mathcal{V}_L^{(t)}, \mathcal{E}_L^{(t)})$ and the receptor graph $\mathcal{G}_R^{(t)} = (\mathcal{V}_R^{(t)}, \mathcal{E}_R^{(t)})$ use atoms as nodes with their respective 3D coordinates as $\mathbf{x}_L^{(t)} \in \mathbb{R}^{N \times 3}$ and $\mathbf{x}_R^{(t)} \in \mathbb{R}^{M \times 3}$, as well as the initial $\psi_h$-dimension roto-translational invariant features $\mathbf{h}_L^{(t)} \in \mathbb{R}^{N \times \psi_h}$ and $\mathbf{h}_R^{(t)} \in \mathbb{R}^{M \times \psi_h}$ (e.g., atom types, electronegativity). Edges include all atom pairs within a distance cutoff of 4~Å. Then the spatial temporal protein sequence is represented as $\{(\mathcal{G}_L^{(t)}, \mathcal{G}_R^{(t)})\}_{t=1}^T$. We denote a vector norm by $x = \|\mathbf{x}\|_2$, and the relative position by $\mathbf{x}_{ij} = \mathbf{x}_i - \mathbf{x}_j$.

### 2.1.2. Prompt-Based Denoising Conformation Generative Task

Generative self-supervised learning is a classic track for unsupervised pre-training.[28–31] It expects to learn effective representations by reconstructing each data point itself. Specifically to drug binding, a good self-supervised learning task should satisfy the following three essential properties. 1) The prediction target is reliable and easy to get. 2) The prediction target should reflect the temporal information within MD trajectories and is relevant to the drug binding. 3) Learned presentations should be diverse and distinguishable.

Guided by these criteria, we present a novel generative prediction task. First, We use a prompt-based pre-training approach to take into account both short-term and long-term temporal dependencies. We set a the prompt embedding $\mathbf{h}_{\text{prompt}} \in \mathbb{R}^{\psi_{\text{prompt}}}$, and append the prompting embedding $\mathbf{h}_{\text{prompt}}^{\Delta t_i}$ to the atomic feature at present timestep $t$ as the model input to achieve the task of predicting the conformation of the next $(t+i)$th step ($i \in \mathbb{Z}^+$), that is, $(\mathbf{x}_L^{(t+i)}, \mathbf{x}_R^{(t+i)})$. This enables our PROTMD approach to not only capture conformational changes between adjacent time steps, but also mine longer-term trajectory information. Then, we perturb the input conformation $(\mathcal{G}_L^{(t)}, \mathcal{G}_R^{(t)})$ with a little random noise at each timestep. It is worth noting that random distortions of the geometry of ligands and receptors at a local energy minimum are almost certainly higher energy configurations.[32] This denoising procedure that maps from a noised molecule to a local energy minimum endows the model with eligibility to learn a map from high energy to low energy, which reveals the binding strength to some extent and maintains exactly the hidden information we expect to encode from MD trajectories. Moreover, since the difference between adjacent conformations or conformations of close steps can be tiny, noise serves to prevent overfitting and makes the pre-training more robust.

*2.1.3. Snapshot Ordering Task*

It has been widely proven that the shape and surface of each conformation carry crucial information for understanding potential molecular interactions.[14] However, previous prompt-based denoising generative task is found on the atom level and may fail to achieve that purpose. It is of necessity for us to formulate a conformation-level self-supervised learning task to catch the global geometric information of conformations.

Inspired by the classic sentence ranking task from NLP,[34,35] we design a snapshot ordering task. To be specific, we require the model to order a set of closely-related conformations as a coherent sub-trajectory, which teaches the model to understand their dependencies from a global perspective.[36]

## 2.2. Equivariant Graph Matching Network

To distinguish the ligand and receptor and also considering the geometric relationship between them in conformational, we refine $E(3)$-EGMN[27] as the molecule encoder network for the PROTMD model. We strictly distinguish the intersections inside and across two graphs $\mathcal{G}_L^{(t)}$ and $\mathcal{G}_R^{(t)}$, respectively as $\mathcal{E}_L^{(t)} \cup \mathcal{E}_R^{(t)}$ and $\mathcal{E}_{LR}^{(t)}$ based on their spatial correlations. It avoids the underutilization of cross-graph edges information (e.g., interatomic distances) due to implicit positional relationships between ligands and receptors. For the protein-ligand at the $t$-th step, we input the set of atom embeddings $\{\mathbf{h}_L^{(t)}, \mathbf{h}_R^{(t)}\}$, and 3D coordinates $\{\mathbf{x}_L^{(t)}, \mathbf{x}_R^{(t)}\}$. Then it outputs a transformation on $\{\mathbf{h}_L^{(t+1)}, \mathbf{h}_R^{(t+1)}\}$ and $\{\mathbf{x}_L^{(t+1)}, \mathbf{x}_R^{(t+1)}\}$, where the latter is exactly the coordinates of the next timeframe. Concisely, $\mathbf{h}_L^{(t+1)}, \mathbf{x}_L^{(t+1)}, \mathbf{h}_R^{(t+1)}, \mathbf{x}_R^{(t+1)} = \text{EGMN}(\mathbf{h}_L^{(t)}, \mathbf{x}_L^{(t)}, \mathbf{h}_R^{(t)}, \mathbf{x}_R^{(t)})$.

## 3. Experiments and Analysis

To thoroughly evaluate the efficiency of the representations learned by our PROTMD, we test its performance in two downstream tasks on the benchmark drug binding dataset with both linear-probing and fine-tuning, and compare it with multiple state-of-the-art methods. The linear-probing updates all model parameters while fine-tuning only updates the last linear layers (e.g., the prediction head).

### 3.1. Dataset and Setup

*3.1.1. Pre-training Data*

In regards to the pre-training data collection, we selected sixty-four protein-ligand pairs in PDBbind and run their MD simulations. Conformations of each protein-ligand pair at a series of time intervals are generated by Amber,[36] a high-performance toolkit widely accepted for molecular simulation. In addition, though the MD simulations are implemented with the whole protein-ligand pair as well as the solvents, we only use the pocket part as the model input instead of the entire protein in PROTMD for the following two major reasons. First, the pocket is the most crucial region that the protein interacts with the ligand, which undergoes the most violent spatial change during the interaction process and can reveal enough information about drug binding. This also explains why we adopt the pocket fraction as the model input for subsequent downstream tasks. Second, the pocket is much smaller and contains far less atoms than the integral protein, so the training speed is significantly faster. To be specific, we locate the pocket as atoms in proteins whose minimum distance to the ligand is shorter than a threshold of 6~Å. More details regarding the experiments and conformation generations are elaborated in Appendix B

*3.1.2. Downstream Data*

Concerning binding affinity prediction, we adopt the PDBbind database,[37,38] a curated database containing protein-ligand complexes from the Protein Data Bank[39] and their corresponding binding strengths. The binding affinity provided by PDBbind is experimentally determined and expressed in molar units of the inhibition constant ($K_i$) or dissociation constant ($K_d$). Accordingly, it is a regression problem. Similar to prior work,[4,14,40] we do not distinguish these constants and predict the negative log-transformed affinity as $pK = -\log(K)$. Besides, we select a 30% sequence identity threshold to limit homologous ligands or proteins and split those complexes into training, test, and validation.

As for the ligand efficacy prediction problem,[4] the dataset is also created from PDB.[40] It contains a curated set of proteins from several families with both "active" and "inactive" state structures. Therefore, it is a traditional binary classification task. There are 527 small molecules with known activating or inactivating function modeled in using the program Glide.[42]

### 3.2. Baselines

We choose wide-ranging popular or state-of-the-art baselines for comparison. Among them, four methods are based on sequences including LSTM,[42] TAPE,[43] ProtTrans,[44] and DeepDTA.[40] They take in pairs of ligand and protein SMILES as the input. While, two other approaches are established on molecular

**Table 1.** Comparison of RMSE, $R_p$, and $R_s$ on PDBbind. The best performance is marked bold and the second best is underlined for clear comparison. Results are reported with the mean and the standard deviation values for three experimental runs.

| Model | # Params | Pre. | Sequence identity (30%) | | |
|---|---|---|---|---|---|
| | | | RMSE | $R_p$ | $R_s$ |
| Sequence-based methods | | | | | |
| DeepDTA [42] | 1.93 M | No | 1.565 ± 0.080 | <u>0.573 ± 0.022</u> | <u>0.574 ± 0.024</u> |
| LSTM [44] | 48.8 M | No | 1.985 ± 0.016 | 0.165 ± 0.006 | 0.152 ± 0.024 |
| TAPE [45] | 93 M | No | 1.890 ± 0.035 | 0.338 ± 0.044 | 0.286 ± 0.124 |
| ProtTrans [46] | 2.4 M | No | 1.544 ± 0.015 | 0.438 ± 0.053 | 0.434 ± 0.058 |
| Surface-based method | | | | | |
| MaSIF [48] | 0.62 M | No | 1.484 ± 0.018 | 0.467 ± 0.020 | 0.455 ± 0.014 |
| Multi-scale methods | | | | | |
| HoloProt [14] | 1.44 M | No | 1.464 ± 0.006 | 0.509 ± 0.002 | 0.500 ± 0.005 |
| Structure-based methods | | | | | |
| 3DCNN [4] | 2.1 M | No | 1.429 ± 0.042 | 0.541 ± 0.029 | 0.532 ± 0.033 |
| IEConv [47] | 5.8 M | No | 1.554 ± 0.016 | 0.414 ± 0.053 | 0.428 ± 0.032 |
| ProtMD (Linear-probing) | 0.01 M [24] | Yes | <u>1.413 ± 0.032</u> | 0.572 ± 0.047 | 0.569 ± 0.051 |
| ProtMD (Fine-tuning) | 5.22 M | Yes | **1.367 ± 0.014** | **0.601 ± 0.036** | **0.587 ± 0.042** |

This corresponds to the number of all trainable parameters (only the predictor).

geometric structures. IEConv[45] designs a convolution operator that considers the primary, secondary, and tertiary structure of proteins and a set of hierarchical pooling operators for multi-scale modeling. 3DCNN[4] is also a competitive 3D method via convolution operations. Additionally, MaSIF[46] takes advantage of protein surfaces. HoloProt[14] introduces a multi-scale construction of protein representations, which connects surface to structure and sequence.

### 3.3. Binding Affinities Predictions

**Table 1** reports the root-mean-squared error (RMSE), the Pearson correlation ($R_p$), and the Spearman correlation ($R_s$) of all baselines. We can observe that our ProtMD not only achieves the lowest RMSE, but also attains the highest Pearson and Spearman correlations compared to these state-of-the-art approaches. This indicates the strong capability and superiority of our pre-training method to learn efficacious representations for the estimation of drug binding.

It is worth noting that ProtMD with linear-probing can realize a RMSE of 1.413, which outperform all baselines. This phenomenon demonstrates the firm correlation between the pre-training and downstream tasks, and deeper insights and exploration are offered in Section 3.6. More importantly, our model is pre-trained only in the trajectories of only sixty-four proteins, but examined in more than 3K proteins. This big gap strongly shows that our ProtMD has great generalization capability. Thus, the need for higher computational expenditure to model the trajectories of a large number of protein-ligand pairs is avoided. On the contrary, pre-training on a small group of binding pairs is adequate, and a more numerical analysis to verify this claim is provided in Section 3.7.

Besides, it can be found that ProtMD with fine-tuning yields a lower RMSE and higher correlations than ProtMD with linear-probing. This proves the necessity of tuning all parameters rather than only the parameters of the output predictor. It can also be demonstrated that structure-based methods generally surpass sequence-based and surface-based approaches.

### 3.4. Ligand Efficacy Prediction

Many proteins switch on or off their function by changing shape. Predicting which shape a drug will favor is thus an important task in drug design. To further demonstrate the validity of our ProtMD, we examine it in the ligand efficacy prediction task. To be explicit, it is formulated as a binary classification task where we predict whether a molecule bound to the structures will be an activator of the protein's function or not. Following Townshend et al.,[4] we only use the regions within a radius of 5.5~Å around the ligand as the model input and adopt a binary cross entropy loss as the loss function. Two metrics are used there: AUROC is the area under the receiver operating characteristic curves, and AUPRC is area under the precision-recall curve. **Table 2** documents the results, which show that our ProtMD can realize the highest values of both AUPRC and AUROC simultaneously.

### 3.5. Generalization Ability to Unknown Structure Pairs

In Table 1, all previous methods are merely evaluated on experimentally known structure pairs. However, in real-world applications, the experimental cocrystal structures are not always accessible. As a remedy, in silico docking algorithms are adopted to predict the binding pose. Therefore, it is of necessity to validate our method on those predicted structures. Here we employ Equibind,[2] a state-of-the-art paradigm, to quickly locate the binding site and the ligand's bound pose and orientation for all test samples in PDBbind. We delete nine unsuccessfully docking pairs, where the distance between the ligand and the receptor is farther than 6 Å (i.e., no pocket exists). **Table 3** reports the results and it can be found that PROTMD still achieves an RMSE as low as 1.474. This shows the generalization ability of our mechanism for unknown structures. Moreover, the decrease of the performance compared to Table 1 mainly comes from the inaccuracy of structure prediction method (i.e., Equibind). It is worth noting that Equibind has a very high ligand RMSE of 8.2 on average, whose predicted structures can significantly mislead our algorithm to forecast the binding properties.

Table 2. Comparison of AUROC and AUPRC on the ligand efficacy prediction task. Results are reported for three experimental runs.

| Metric | 3DCNN [4] | 3DGCN [4] | Cormorant [48] | DeepDTA [41] | PROTMD (Linear-probing) | PROTMD (Fine-tuning) |
|---|---|---|---|---|---|---|
| AUROC | 0.589 ± 0.020 | 0.681 ± 0.062 | 0.663 ± 0.100 | 0.696 ± 0.021 | 0.548 ± 0.112 | 0.742 ± 0.039 |
| AUPRC | 0.483 ± 0.037 | 0.598 ± 0.135 | 0.551 ± 0.121 | 0.550 ± 0.024 | 0.516 ± 0.138 | 0.724 ± 0.041 |

Table 3. Comparison of RMSE, $R_p$, and $R_s$ on predicted structures.

| Model | Sequence Identity (30%) | | |
|---|---|---|---|
| | RMSE | $R_p$ | $R_s$ |
| No Pretrain | 1.668 ± 0.048 | 0.447 ± 0.039 | 0.499 ± 0.041 |
| Fine-tuning | 1.474 ± 0.030 | 0.517 ± 0.038 | 0.508 ± 0.044 |

### 3.6. Why Does PROTMD Work?

Ligands and receptors explore binding sites during the dynamic interaction process. Our experiments also firmly demonstrate the effectiveness of PROTMD capable of capturing such time-dependent dynamic binding information, but why self-supervised learning on MD trajectories is beneficial for downstream tasks still requires further analysis. In other words, we desire to quantitatively understand the correlation between MD simulations and drug binding-related properties such as the binding affinity. Notably, our design of PROTMD leads to transformations on both features and 3D coordinates, while the pre-training stage and the fine-tuning stage utilize different outputs to calculate the losses. On the one hand, the 3D coordinates is used in the self-supervised pre-training stage and correspond to the coordinates of future timeframe. On the other hand, the updated features is used in the fine-tuning stage and participate in acquiring the properties of the ligand-receptor pair. Therefore, it is motivating for us to explore the relationship between the outcome 3D coordinates and properties like binding affinities for our PROTMD before fine-tuning or linear-probing. To be specific, we compute the average predicted space shift between the input and output 3D coordinates for the ligand-receptor pair as $\Delta x_{LR} = \frac{\|x_{LR}^{out} - x_{LR}^{in}\|_2^2}{N+M}$. $\Delta x_{LR}$ corresponds to the magnitude of the movement in the 3D space predicted by our PROTMD. Then we investigate the relation between this predicted spatial motion and the ground truth properties.

**Figure 3** exhibits the results. We perform a linear least-squares regression and attain that $y = -65.90 \Delta x_{LR} + 7.91$ with $R^2 =$

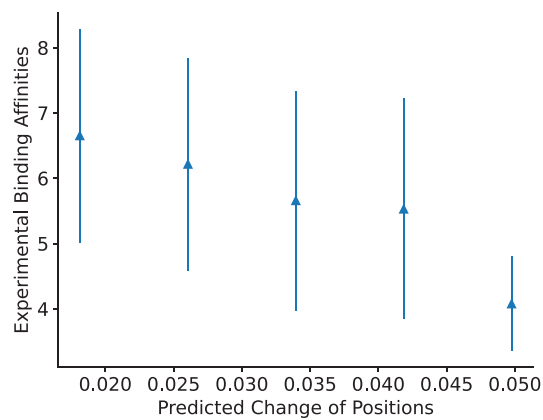

(a) Ligand Binding Affinity Prediction

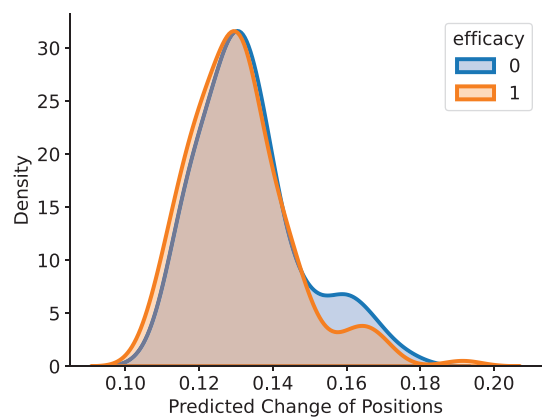

(b) Ligand Efficacy Prediction

**Figure 3.** Visualization of predicted spatial change and the corresponding drug binding-related properties. a) The error plot of the relationship between the predicted positional change and the binding affinities. b) The density plot of the predicted positional change and the corresponding efficacy.

−0.2601. It can be observed that for ligand binding affinity task, the change of spatial positions is highly associated with the binding affinities. For instance, if PROTMD forecasts a smaller change in the coordinates of the binding site, then this pair is more likely to have better binding interactions. This claim aligns with the discovery in Guterres and Im[49] that ligands with good initial binding modes tend to stay stable during MD simulations. Similarly, in the ligand efficacy prediction task, the average change is also concerned with the activity of protein-ligand pairs. To be concrete, for ligands with efficacy = 0, the predicted change of positions is 0.1335 ± 0.0144 (mean ± std). For ligands with efficacy = 1, the predicted change of positions is 0.1309 ± 0.0141. This indicates that more efficacious ligands have a slightly smaller positional change. We also calculate the Davis Bouldin (DB) index[50] to measure the separation of the two efficacy clusters, and it achieves a DB index as low as 8.09. All the above analysis confirms that even though the pre-training stage and the fine-tuning stage employ different parts of the output of our PROTMD for backpropagation to update the model parameters (i.e., 3D coordinates for the pre-training stage and atomic features for the fine-tuning stage), PROTMD can properly capture the inherent link between 3D positions and their corresponding drug binding-related properties.

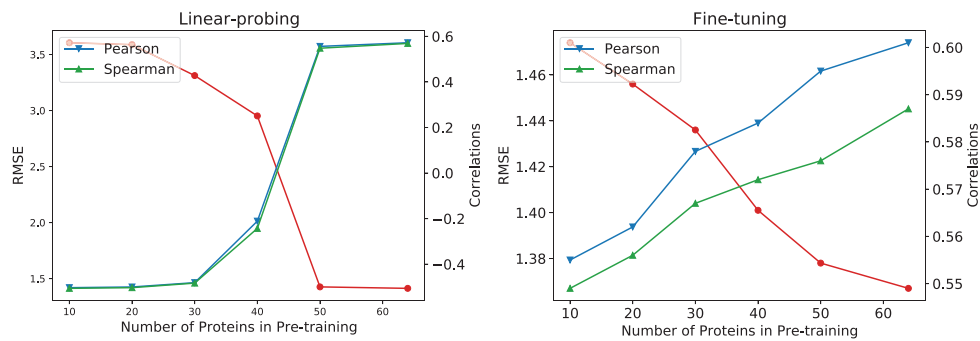

**Figure 4.** Ablation study on the number of protein-ligand pairs used in the pre-training stage, where the red line denotes the RMSE, and the blue and green lines denote the Pearson and Spearman correlations, respectively. The left and right figures correspond to different strategies of linear-probing and fine-tuning, respectively.

### 3.7. How Many Trajectories Do PROTMD Need?

As previously announced, even though our pre-training dataset only contains the trajectories of approximately sixty protein-ligand pairs, our model realizes unexpected generalization to all 3K samples in PDBbind. Thus, we take a further step to investigate the influence of the number of pre-training samples over the performance in downstream tasks. As shown in **Figure 4**, when there are only few proteins (e.g., 10 or 20 proteins), PROTMD with linear-probing performs badly. Its Pearson and Spearman correlations are negative. Nevertheless, when the number of proteins exceeds 50, the benefit for linear-probing is negligible. On the other hand, the improvement for fine-tuning persistently augments along with the increase of the number of pre-training samples. However, the increments of RMSE, Pearson, and Spearman correlations also become smaller when the number of proteins used in the pre-training increases. Thus, it would be a contribution of providing a more beneficial pre-training database with a larger size and we leave it for future work to produce trajectories of more protein-ligand pairs.

### 3.8. Ablation Studies and Visualization

We also conduct extensive experiments to exam the effects of each component in our PROTMD in the ligand binding affinity prediction. First, we analyze if the performance of PROTMD with two self-supervised learning tasks outperforms its isolated components, that is, when using only one task for pre-training. The second ablation axis analyzes the benefits of the noise injection and the prompt for the generative self-supervised learning.

As displayed in **Table 4**, the results clearly show that both self-supervised learning tasks contribute to the efficacy of learned representations. We further observe that both the noise trick and the prompt are more useful for linear-probing than fine-tuning, leading to a decrease of 0.068 in RMSE, an increase of 0.034 in the Pearson correlation, and an increase of 0.039 in the Spearman correlation. This phenomenon supports our statement that noise can effectively increase the difficulty of pre-training task and prevent overfitting. Interestingly, fine-tuning the pre-trained model with a noiseless and non-prompt generative task can already realize very outstanding performance, which is even better than any sort of model with linear-probing.

To intuitively observe the representations that our self-supervised learning tasks have learned, we envision the representations by mapping them to the 2D space by the principal component analysis (PCA) and TSNE[51] algorithms in **Figure 5**. Remarkably, even without any label information, the representations learned from self-supervised learning follow some kind of pattern that is strongly related to the drug binding-related properties. For the efficacy binary classification problem, it has a DB index of 4.05. This demonstrates that our designed self-supervised learning tasks are an appropriate way to excavate structural and dynamical properties of molecular systems and comprehend the mechanism of physiochemical processes within the MD trajectory. It also aligns with the previous analysis that our PROTMD can achieve extraordinary performance in binding affinity prediction via linear-probing.

## 4. Related Work

### 4.1. Protein-Ligand Modeling

With increasing availability of sequence and structure data, the area of protein representation learning has developed rapidly.[52] Free energy-based simulations and DL-based scoring functions are two major computational methods for the binding affinity prediction.[53] The latter is completely data-driven and can fast screen a vast number of compounds, attaching increasing interests.

1D amino acid sequences continue to be the simplest and most abundant source of protein data, resulting in various methods[40,53,54] that borrow ideas from the area of NLP. Beyond that, previous methods ignore the spatial complexity of proteins and it has been proven that the exploitation of their 3D structures leads to improved performance,[4,55] which is further supported by the revolution in protein structure prediction.[51] Some utilize 3D grids to capture the spatial distribution of the properties within molecular conformers, where 3DCNN[3,56] have been the method of choice. Other studies use 3D voxel-based surface representations as inputs to 3DCNN[57,58] for the prediction of protein-ligand binding sites.[59] Apart from them, a protein structure can be naturally represented as a proximity graph over amino acid nodes, and a number of graph neural networks (GNNs) are been proposed.[60] For instance, structured transformer[61] is designed for protein design, which encodes 3D geometry through relative orientations, and GraphQA[62] is introduced to solve model quality assessment. IEConv[45] introduces a graph convolution layer to incorporate both intrinsic and extrinsic distances between nodes. GVP[63,64] extends standard dense layers to operate on collections of Euclidean vectors. However, none of those DL approaches consider a temporal perspective and take advantage of molecular dynamics simulations to describe the joint flexibility of proteins and ligands.[65]

**Table 4.** Ablation study on ProtMD. The DG (naive) stand for the generative task without noise and prompt, and the SO stand for the snapshot ordering task. Results are documented with the mean and standard deviation for three runs.

|  | DG (naive) | Noise | Prompt | SO | RMSE | $R_p$ | $R_s$ |
|---|---|---|---|---|---|---|---|
| No pre-train | - | - | - | - | 1.541 ± 0.030 | 0.542 ± 0.048 | 0.527 ± 0.041 |
| Linear-probing | ✓ | - | - | - | 1.498 ± 0.030 | 0.533 ± 0.047 | 0.519 ± 0.039 |
|  | ✓ | ✓ | - | - | 1.449 ± 0.028 | 0.536 ± 0.041 | 0.523 ± 0.035 |
|  | ✓ | ✓ | ✓ | - | 1.430 ± 0.021 | 0.567 ± 0.036 | 0.558 ± 0.042 |
|  | ✓ | ✓ | ✓ | ✓ | 1.413 ± 0.020 | 0.572 ± 0.035 | 0.569 ± 0.048 |
| Fine-tuning | ✓ | - | - | - | 1.403 ± 0.034 | 0.582 ± 0.038 | 0.561 ± 0.039 |
|  | ✓ | ✓ | - | - | 1.386 ± 0.049 | 0.592 ± 0.044 | 0.578 ± 0.032 |
|  | ✓ | ✓ | ✓ | - | 1.372 ± 0.027 | 0.600 ± 0.045 | 0.584 ± 0.033 |
|  | ✓ | ✓ | ✓ | ✓ | 1.367 ± 0.024 | 0.601 ± 0.042 | 0.587 ± 0.030 |

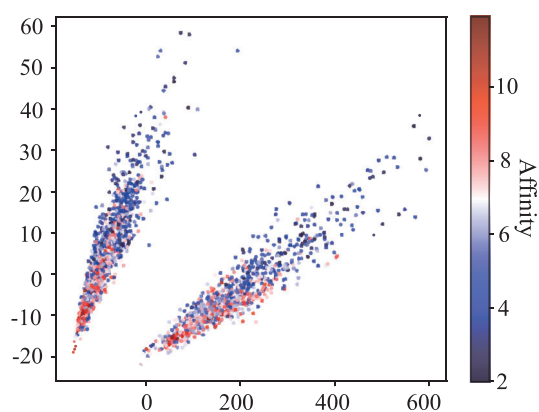

(a) Ligand Binding Affinity Prediction

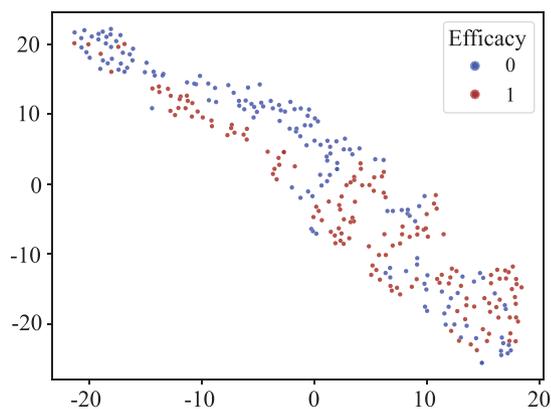

(b) Ligand Efficacy Prediction

**Figure 5.** Dimensionality reduction of the protein-ligand representations learned from our self-supervised learning tasks. The sub-figure (a) is drawn by PCA and the color is based on the strength of corresponding binding affinities. The sub-figure (b) is drawn by TSNE[51] and the color corresponds to different types of efficacy.

### 4.2. Protein Self-Supervised Learning

The self-supervised learning gains great progress on NLP tasks,[67] and inspired by that, many methods have been expanded to the biological area.[68,69] Regarding representing proteins, most preceding studies focus on pre-training on unlabeled amino acid sequences because of their abundance.[43–45,70,71] For instance, TAPE[44] uses the masked-token mechanism to pre-train the model and achieves good performance on several sequence-based prediction tasks. However, due to the fact that protein functions are heavily governed by their folded structures, more and more attention has been draw to leverage the full spatial complexity of proteins. Hermosilla and Ropinski[72] uses contrastive learning for representation learning of 3D protein structures from the perspective of sub-structures. Apart from that, Zhang et al.[73] combines a multi-view contrastive learning and a self-prediction learning to encode geometric features of proteins. Then these semantic representations learned from self-supervised learning are utilized for downstream tasks including structure classification,[46] and function prediction.[61] Nevertheless, no preceding research excavate the potential of pre-training on this sort of spatial-temporal data and enable the model to understand the flexibility of proteins, partly because of the high expenditure to run MD simulations.

## 5. Conclusion

Biological discoveries demonstrates that the flexibility of both the receptor and ligand is deterministic in deciding strength of drug binding, and MD simulations conventionally shoulder the responsibility to depict this dynamical process. In this work, in order to employ the time-dependent information of flexibility inside MD trajectories, we introduce a simple yet effective pre-training paradigm from both spatial and temporal perspectives for protein representation learning called PROTMD. It consists of two categories of self-supervised learning tasks: one is the atom-level prompt-based denoising generative task and the other is the conformation-level snapshot ordering task. And the improved EGMN as the backbone of protMD, to joint transformation of features and three-dimensional coordinates to achieve information transfer within and between graphs. We then linear probe and fine-tune the pre-trained models in the downstream drug binding prediction. Extensive experiments verify its effectiveness and ablation studies demonstrate the necessity of each component of our proposed PROTMD.

## 6. Experimental Section

*Future Conformation Prediction*: The conformation $(\mathcal{G}_L^{(t+1)}, \mathcal{G}_R^{(t+1)})$ of the next timeframe was used as the target and models were required to forecast this prospective position. The objective was to maximize the likelihood as

$$L_1 = \sum_{t=k}^{T} \log P\left(\left(\mathbf{x}_L^{(t+1)}, \mathbf{x}_R^{(t+1)}\right) \mid \left\{\left(\mathcal{G}_L^{(i)}, \mathcal{G}_R^{(i)}\right)\right\}_{i=1}^{t}; \theta\right) \quad (1)$$

where $k$ is the size of context window, and the conditional probability $P$ is modeled via the encoder $f_\theta$. Conventionally, several frameworks assumed the Markov property on biomolecular conformational dynamics[73,74] for ease of representation, that is, $P((\mathbf{x}_L^{(t+1)}, \mathbf{x}_R^{(t+1)}) \mid \{(\mathcal{G}_L^{(i)}, \mathcal{G}_R^{(i)})\}_{i=1}^{t}) = P((\mathbf{x}_L^{(t+1)}, \mathbf{x}_R^{(t+1)}) \mid \mathcal{G}_L^{(t)}, \mathcal{G}_R^{(t)})$. This rule was obeyed and therefore the length of context window was set as $k = 1$. As a consequence, the goal becomes:

$$L_1 = \log P\left(\left(\mathbf{x}_L^{(t+1)}, \mathbf{x}_R^{(t+1)}\right) \mid \left\{\mathcal{G}_L^{(t)}, \mathcal{G}_R^{(t)}\right\}; \theta\right) \quad (2)$$

*Time-Series Prompting for Motion Prediction*: The previous formation of generative pre-training only guaranteed the model to capture the change of conformations between adjacent timesteps. There exist other sorts of information that could only be excavated from long-term trajectories. In other words, in addition to the prediction of $(\mathbf{x}_L^{(t+1)}, \mathbf{x}_R^{(t+1)})$, it was more reasonable to include $(\mathbf{x}_L^{(t+i)}, \mathbf{x}_R^{(t+i)})$ ($i > 1$) as the prediction target. A potential solution was to rely on the multi-task learning, which allowed the model to simultaneously output conformations of different time intervals. Nevertheless, with multi-task learning, the model intended to use a generalized representation and neglect the nuance between long-term and short-term trajectories.

Prompt tuning, with the emerging of GPT-3,[76] had gradually become a standard genre for pre-trained model tuning. By designing, generating, and searching discrete or continuous prompts,[77,78] the gap between pre-training and fine-tuning was bridged, and the computational cost on fine-tuning the tremendous amounts of parameters was reduced. Motivated by these attractive benefits of prompts, a novel prompt-based pre-training approach was proposed to concern both short-term and long-term temporal dependencies. By searching explicitly, the prompt embedding $\mathbf{h}_{\text{prompt}} \in \mathbb{R}^{\psi_{\text{prompt}}}$ was concatenated to the atomic features of both ligand and receptor $\mathbf{h}_L^{(t)}$ and $\mathbf{h}_R^{(t)}$, respectively. This prompt served as an indicator for the model to predict the new conformation after a certain period. To be specific, suppose there was a set of pre-defined time intervals $\{\Delta t_1, \Delta t_2, \ldots\}$, each time interval $\Delta t_i \in [T-1]$ had a corresponding learnable prompt embedding $\mathbf{h}_{\text{prompt}}^{\Delta t_i}$. Then if the model was expected to forecast the conformation after $\Delta t_i$, that is, $(\mathbf{x}_L^{(t+\Delta t_i)}, \mathbf{x}_R^{(t+\Delta t_i)})$, the prompting embedding $\mathbf{h}_{\text{prompt}}^{\Delta t_i}$ was appended to the atomic feature at present timestep $t$ as the model input. Afterward, the dimension of initial $\psi_h$-dimension roto-translational invariant feature becomes $\psi_h + \psi_{\text{prompt}}$. During fine-tuning, a new prompt embedding $\mathbf{h}_{\text{prompt}}^* \in \mathbb{R}^{\psi_{\text{prompt}}}$ is assigned to help model differentiate the self-supervised learning pre-training and the specific downstream task.

*Noise Perturbation*: Moreover, the input conformation $(\mathcal{G}_L^{(t)}, \mathcal{G}_R^{(t)})$ was perturbed with a little noise at each timestep. This operation of perturbation had both theoretical and empirical supports. As demonstrated by Wu, Zhang, Jin, Jiang, and Li,[79] the denoising diffusion architecture[80–82] had a strong connectivity with the enhanced sampling method in MD,[83–86] where energy was injected into the microscopic system to smooth biomolecular potential energy surface and decrease energy barriers. Besides, it had been shown in Godwin et al.[33] that the simple noise regularization could be an effective way to address oversmoothing.[87] A noise correction target could be added to prevent oversmoothing by enforcing diversity in the last few layers of the GNN, which was implemented with an auxiliary denoising autoencoder loss. In addition, it was argued that as the difference between neighboring snapshots was small, the perturbation played a critical part in preventing overfitting and improve generalization.[88]

Toward this goal, the corrupted conformation is defined as $(\tilde{\mathcal{G}}_L^{(t)}, \tilde{\mathcal{G}}_R^{(t)})$. The coordinates $\tilde{\mathbf{x}}_L^{(t)} = \mathbf{x}_L^{(t)} + \sigma^{(t)}$ and $\tilde{\mathbf{x}}_R^{(t)} = \mathbf{x}_R^{(t)} + \sigma^{(t)}$ is constructed by adding a noise, which is drawn from a normal distribution as $\sigma^{(t)} \sim \mathcal{N}(0, \sigma^2)$ and $\sigma$ is a pre-defined hyperparamter to control the magnitude of perturbation. Remarkably, the translational invariant features of the ligand and the receptor $\mathbf{h}_L^{(t)}$ and $\mathbf{h}_R^{(t)}$ can be either perturbed or not.

*Snapshot Ordering Pre-training*: A set of $n$ conformations with the order $\mathbf{t} = \{t_i\}_{i=1}^{n}$ could be described as $\{(\mathcal{G}_L^{(t_i)}, \mathcal{G}_R^{(t_i)})\}_{i=1}^{n}$. The goal is to find the correct order $\mathbf{t}^*$ for them, with which the whole sub-trajectory has the greatest coherence probability as:

$$P\left(\mathbf{t}^* \mid \left\{\left(\mathcal{G}_L^{(t_i)}, \mathcal{G}_R^{(t_i)}\right)\right\}_{i=1}^{n}; \theta\right) \geq P\left(\mathbf{t} \mid \left\{\left(\mathcal{G}_L^{(t_i)}, \mathcal{G}_R^{(t_i)}\right)\right\}_{i=1}^{n}; \theta\right), \forall \mathbf{t} \in \Upsilon, \quad (3)$$

where $\mathbf{t}$ indicates any order of these conformations and $\Upsilon$ denotes the set of all possible orders. A series of approaches had been invented to tackle this rearranging problem.[89–92] There the topological sort method was selected,[93,94] a standard algorithm for linear ordering of the vertices of a directed graph. Precisely, we have $C_n$ set of constraints for this sub-trajectory. These constraints $C_n$ represent the relative ordering between every pair of conformations in $\{(\mathcal{G}_L^{(t_i)}, \mathcal{G}_R^{(t_i)})\}_{i=1}^{n}$. Hence, we have $|C_n| = \binom{v_j}{2}$, and constraints $C_n$ are learned using a multi-perceptron-layer (MLP) classifier. Notably, if we make $n = 2$, the snapshot ordering task is deformed to the next sentence prediction (NSP),[21] which is criticized as a weak task for its comparison of similarity.[95]

*Equivariant Graph Matching Neural Network*: Equivariance is ubiquitous in deep learning for microscopic systems. This was because the physical law controlling the dynamics of atoms stays the same regardless of the rotation and translation of biomolecules.[95] Thus, it was essential to incorporate such inductive bias symmetry into model parameterization for modeling 3D geometry and achieving better generalization capacity.[96–99] Moreover, there were two distinct graphs in the circumstance, where the ligand graph $\mathcal{G}_L^{(t)}$ was much smaller than the receptor graph $\mathcal{G}_R^{(t)}$. To simply combine these two graphs together would confound the network and better representation capability was prohibited due to the nondiscrimination of small molecules and proteins. Therefore, it was of great need to make this model aware of the distinction between these two components.

To satisfy these requirements, inspirations were drawn from recent models[97,100,101] and a variant of *E(3)-Equivariant Graph Matching Network* (EGMN)[27] was employed as the molecule encoder. Notably, this refined EGMN has several key distinctions from prior work. On the one hand, receptor structures in Ganea et al.[27] were rigid and did not move during the whole binding process. Apart from that, since the positional relationship between ligand and receptor in their setting was implicit, they were unable to fully exploit the information within cross-graph edges (e.g., inter-atomic distances). As an alternative, they aggregated *intra*-messages assuming fully-connected edges $\mathcal{E}_{LR}^{(t)}$ between $\mathcal{G}_L^{(t)}$ and $\mathcal{G}_R^{(t)}$. On contrast, the intersections inside and across two graphs $\mathcal{G}_L^{(t)}$ and $\mathcal{G}_R^{(t)}$ were strictly distinguished respectively as $\mathcal{E}_L^{(t)} \cup \mathcal{E}_R^{(t)}$ and $\mathcal{E}_{LR}^{(t)}$ based on their spatial correlations.

Concisely, the cross-graph edges $\mathcal{E}_{LR}^{(t)}$ based on their atomic pairwise distances in addition to the internal edges of $\mathcal{G}_L^{(t)}$ and $\mathcal{G}_R^{(t)}$ were constructed. Then the layer of EGMN was formally defined as the following:

$$\mathbf{m}_{j \to i} = \phi_e\left(\mathbf{h}_i^{(t),l}, \mathbf{h}_j^{(t),l}, x_{ij}^{(t),l}\right), \forall e_{ij} \in \mathcal{E}_L^{(t)} \cup \mathcal{E}_R^{(t)}, \quad (4)$$

$$\mu_{j \to i} = a_{j \to i} \mathbf{h}_j^{(t),l} \cdot \phi_d\left(x_{ij}^{(t),l}\right), \forall e_{ij} \in \mathcal{E}_{LR}^{(t)}, \quad (5)$$

$$\mathbf{x}_i^{(t),l+1} = \mathbf{x}_i^{(t),l} + \left(\mathbf{x}_i^{(t),l} - \mathbf{x}_j^{(t),l}\right)\phi_x(i, j^*) \quad (6)$$

$$\mathbf{h}_i^{(t),l+1} = \phi_h\left(\mathbf{h}_i^{(t),l}, \sum_j \mathbf{m}_{j \to i}, \sum_{j'} \mu_{j' \to i}\right), \quad (7)$$

where $\phi_e$ is the edge operation, and $\phi_h$ denotes the node operation that aggregates the *intra*-graph messages $\mathbf{m}_i = \sum_j \mathbf{m}_{j \to i}$ and cross-graph message $\mu_i = \sum_{j'} \mu_{j' \to i}$ as well as the node embeddings $\mathbf{h}_i^{(t),l}$ to acquire the updated node embedding $\mathbf{h}_i^{(t),l+1}$. $\phi_x$ varies according to whether the edge $e_{ij}$ is *intra*-graph or cross-graph. Particularly, $\phi_x = \phi_m(\mathbf{m}_{i \to j})$ if $e_{ij} \in \mathcal{E}_L^{(t)} \cup \mathcal{E}_R^{(t)}$. Otherwise, $\phi_x = \phi_\mu(\mu_{i \to j})$ when $e_{ij} \in \mathcal{E}_{LR}^{(t)}$, where $\phi_m$ and $\phi_\mu$ are two different functions to cope with different kinds of messages. It takes as input the edge embedding $\mathbf{m}_{i \to j}$ or $\mu_{i \to j}$ as the weight to sum all relative distance $\mathbf{x}_i^{(t),l} - \mathbf{x}_j^{(t),l}$ and output the renewed coordinates $\mathbf{x}_i^{(t),l+1}$. $\phi_d$ operates on the inter-atomic distances $x_{ij}^{(t),l}$. $a_{j \to i}$ is an attention weight with trainable MLPs $\phi^q$ and $\phi^k$, and takes the following form as:

$$a_{j \to i} = \frac{\exp\left(\left\langle \phi^q\left(\mathbf{h}_i^{(t),l}\right), \phi^k\left(\mathbf{h}_j^{(t),l}\right)\right\rangle\right)}{\sum_{j'} \exp\left(\left\langle \phi^q\left(\mathbf{h}_i^{(t),l}\right), \phi^k\left(\mathbf{h}_{j'}^{(t),l}\right)\right\rangle\right)} \quad (8)$$

Specifically, the *l*th layer of our encoder ($l \in [L]$) took as input the set of atom embeddings $\{\mathbf{h}_L^{(t),l}, \mathbf{h}_R^{(t),l}\}$, and 3D coordinates $\{\mathbf{x}_L^{(t),l}, \mathbf{x}_R^{(t),l}\}$. Then it outputs a transformation on $\{\mathbf{h}_L^{(t+1),l}, \mathbf{h}_R^{(t+1),l}\}$ and $\{\mathbf{x}_L^{(t+1),l}, \mathbf{x}_R^{(t+1),l}\}$, where the latter is exactly the coordinates of the next timeframe. Concisely, $\mathbf{h}_L^{(t+1),l+1}, \mathbf{x}_L^{(t+1),l+1}, \mathbf{h}_R^{(t+1),l+1}, \mathbf{x}_R^{(t+1),l+1} = \text{EGMN}(\mathbf{h}_L^{(t)}, \mathbf{x}_L^{(t)}, \mathbf{h}_R^{(t)}, \mathbf{x}_R^{(t)})$.

*Fine-Tuning and Linear-Probing*: When fine-tuning, since there was only one snapshot for each protein-ligand pair, the temporal superscript was omitted. Then pool $\mathbf{h}_i$ was averaged across all atoms in both protein and ligand to extract a $\psi_h$-dimensional vector of features per example as $\mathbf{H} = \text{Pool}(\{\mathbf{h}_i\}_{i=1}^N) \in \mathbb{R}^{\psi_h}$. A projection was expected to learn from $\mathbf{H}$ to the target property such as the binding affinities and the ligand efficacy. We use a root-mean-squared-error (RMSE) loss and a binary cross-entropy loss to supervise the training for them, respectively.

Extracting features for linear probing follows a similar procedure to fine-tuning, except that those features are fixed and the encoder does not participate in the backpropagation.

# Appendix A: Molecular Dynamics Simulations

## A.1. More Introductions about MD Simulations

Ab initio MD techniques have long and extensively been used to investigate structural and dynamical properties of a wide variety of molecular systems and understand the mechanism of physiochemical processes.[102–104] It substantially accelerates the studies to observe biomolecular process in action, particularly important functional processes such as ligand binding,[105] ligand- or voltage-induced conformational change,[106] protein folding,[107] or membrane transport.[108,109]

The most basic and intuitive application of MD is to assess the mobility or flexibility of various regions of a biomolecule. Instead of yielding an average structure by experimental structure determination methods including X-ray crystallography and cryo-EM, MD allows researchers to quantify how much various regions of the molecule move at equilibrium and what types of structural fluctuations they undergo, which is critical for protein function and ligand binding.[110–112] To be explicit, on the one hand, simulations of the full ligand-binding process can reveal the binding site and pose of a ligand.[105,113–115] On the other hand, at a quantitative level, simulation-based methods provide essentially more accurate estimates of ligand binding affinities (free energies) than other computational approaches such as docking.[116]

## A.2. MD Simulations in Experiments

We use the pdb4amber program[37] to prepare the original PDB file downloads from the RCSB. The tags of all non-standard residues are sorted and the ones with the fewest occurrences are selected to be merged with the protein, and obtain the complex files. These files are manually inspected by a pharmaceutical expert to determine whether they are suitable as protein-ligand complex models.

It takes an RTX3080 GPU ≈20 h to run 100 ns per protein-ligand complex with the periodic boundary condition in the NPT ensemble. The detailed steps are described as the following:

1) The solvated system is conducted 5000 steps of minimization by specifying MAXCYC = 5000, sander will use the steepest descent algorithm for the first NCYC = 2500 steps before switching to the conjugate gradient algorithm for the remaining (MAXCYC - NCYC).
2) The NVT simulation is heated gradually from 0 to 303.15 K in the NVT ensemble during a period of 500 ps. The heated system is equilibrated in the NPT ensemble during a period of 1 ns.
3) Production simulation at the temperature of 303.15 K and the pressure of 1 atmospheric pressure (atm). The SHAKE algorithm is used to constrain all covalent bonds involving hydrogens and does not calculate the forces of bonds containing hydrogen. Finally, a 100 ns production simulation is performed and the structure snapshots are collected every 1 ps.

The names of pairs used in our PROTMD are listed as follows: 1TBF, 1TXI, 1ZKL, 1ZP5, 2E1Q, 2GH5, 2I0E, 2JED, 2JIF, 2NO6, 2Z5X, 3B6H, 3D4S, 3DPK, 3FVO, 3I8V, 3IAR, 3IW4, 3JZB, 3LW0, 3OLL, 3QXM, 3ROD, 3TKM, 3W2T, 4DJH, 4IAQ, 4IB4, 4MUW, 4NQD, 4PXZ, 4QTB, 4RWD, 4UDA, 4UXQ, 5AFJ, 5AX3, 5C37, 5CGD, 5DIQ, 5DSG, 5DYY, 5EDU, 6R4V, 6SSQ, 6WV3, 6X40, 7AOS, 7AYM, 7B0V, 7BR3, 7BVQ, 7BW1, 7C7S, 7CMV, 7CX3, 7DFW, 7DHL, 7EO4, 7JVP, 7JVR, 7LRC, 7VNR. The data will be realized once our paper gets accepted.

# Appendix B: Experimental Details

## B.1. EGMN Architecture

For each ligand-receptor pair, there are 10 000 timestep (i.e., $T = 10\~000$). The dimension of roto-translational invariant features $\psi_h$ and the prompt

**Table B1.** The training hyper-parameters.

| Hyper-parameter | Description | Range |
| --- | --- | --- |
| lr | The initial learning rate of ReduceLROnPlateau learning rate scheduler. | [$1e-4, 1e-5$] |
| min_lr | The minimum learning rate of ReduceLROnPlateau learning rate scheduler. | [$5e-6.5e-7$] |
| noise | The magnitude of noise injected into the input conformation during the pre-training stage. | [$1e-5, 1e-3, 1e-1, 1$] |

embeddings $\psi_{prompt}$ are both 128. The pre-defined time intervals are set as [1, 5, 10].

We use a 6-layer EGMN with a hidden dimensionality of 256, the twice of the input feature dimension. We adopt the coordinate normalization and a sum pooling method. The coordinate clamping value is set as 2. A dropout rate of 0.15 is used for any layer. The maximum number of nodes for the input graph is set as 10 000.

### B.2. Downstream Dataset Splits

According to Townshend et al.,[4] the split based on a 30% sequence identity threshold leads to training, validation, and test sets of size 3507, 466, and 490, respectively. In regards to the ligand efficacy prediction, complex pairs is splitted by protein target, the training, validation, and test sets have 608, 208, and 208 samples, respectively.

### B.3. Training Details

We use Pytorch[117] to implement EGMN and a default random seed of 1234. On the pre-training stage, utilize the distributed training with 4 V100 GPUs and a batch size of 32 for each GPU. An Adam[118] optimizer is used and a ReduceLROnPlateau scheduler is enforced to adjust it with a factor of 0.6 and a patience of 10. The initial learning rate is $1 \times \sim 10^{-4}$, and we apply no weight decay there. Each model is trained with 200 epochs. We split the trajectory of each protein-ligand pair into training and validation with a ratio of 9:1, and save the best model based on its performance on the validation set.

On the downstream fine-tuning and linear-probing stage, the number of GPUs, the configurations of optimizer and scheduler keep the same. The batch size is 64. Following Townshend et al.,[4] we also only use the pocket position and the ligand as the model input. Only atoms within a distance of 6 Å from the ligand are used and the number of atoms in total (ligand + protein) is limited to no more than 600. We perform a hyperparamter sweep in **Table B1** for different pre-training models and different strategies of linear-probing and fine-tuning.

### B.4. Baselines

For protein-ligand binding affinity prediction, we use the reported values from Somnath et al.[14] and Townshend et al.[4] As for the model size, we use all available reported numbers from Somnath et al.[14] For 3DCNN, we download the code from the official repository[119] and compute the model parameters. For ligand efficacy predicition, we use the baseline values from Townshend et al.[4]

### B.5. Visualization Details

For the visualization in the main text, we try several different approaches for dimension reduction including PCA, TSNE, uniform manifold approximation and projection (UMAP), and linear discriminant analysis (LDA). The results turn out that PCA and TSNE perform the best for our two downstream tasks, respectively. Since the efficiency of dimension reduction technique is out of the scope of this paper, we believe it will not cause any problem if different dimension reduction methods are applied to visualize the high-dimension representations.

To be concise, we apply the following setting of the TSNE algorithm to the presentations of protein-ligand pairs **H** for the dimension reduction in the ligand efficacy prediction. Concretely, the maximal iteration is 10 000. The perplexity is 30.0. The learning rate is 200. The early exaggeration is 12.0. The angle is 0.5. Finally, the t-SNE reduces the outputs of EGMN into the 2D representations, which then are plotted as 2D images. Additionally, we adopt PCA to reduce the dimension of EGMN's outcome using the popular *sklearn.decomposition* package. We use the default setting with an automatic singular value decomposition (SVD) solver. Thus, it uses the LAPACK implementation of the full SVD or a randomized truncated SVD, depending on the shape of the input data and the number of components to extract.

## Acknowledgements

F.W., S.J., and Y.J., contributed equally to this work. This work was supported in part by the National key research and development program (Grant No. 2021YFA1301603) and National Natural Science Foundation of China (No. U21A20427).

## Conflict of Interest

Y.J., X.J., B.T and Z.N. are the employees of the MindRank AI Ltd. The other authors declare no conflict of interest.

## Author Contributions

F.W., S.Z.L., and S.J led the research. F.W., S.Z.L., S.J. and Y.J. contributed technical ideas. F.W., S.J., X.J. and Q.Z. developed the proposed method. F.W., S.J., Z.N., X.Z and X.L. performed analysis. S.Z.L., X.L., X.Z., Q.Z. and Z.N. provided evaluation and suggestions. All authors contributed to the manuscript.

## Data Availability Statement

The data that support the findings of this study are openly available in PDBbind database; Protein Data Bank at http://www.pdbbind.org.cn/; https://www.rcsb.org, reference number 69. These data were derived from the following resources available in the public domain: PDBbind database; Protein Data Bank, http://www.pdbbind.org.cn/; Protein Data Bank, https://www.rcsb.org/.